\documentclass[
reprint,           
superscriptaddress,
amsmath,           
amssymb,           
aps,               
prd,               
notitlepage,       
longbibliography,  
floatfix,          
nofootinbib,
]{revtex4-1}

\usepackage{tensor}     
\usepackage{float}
\usepackage[caption = false]{subfig} 
\usepackage[final]{graphicx}   
\usepackage[
colorlinks=true,        
citecolor=blue,         
linkcolor=blue,         
urlcolor=blue           
]{hyperref}             
\usepackage{bm}         
\usepackage{xcolor}     
\usepackage{lipsum}
\usepackage{color}      
\usepackage[utf8]{inputenc} 
\usepackage[section]{placeins} 

\newcommand{\nc}{\newcommand*} 

\nc{\al}{\alpha}
\nc{\s}{\sigma}
\nc{\dt}{\delta}
\nc{\Dt}{\Delta}
\nc{\Ld}{\Lambda}
\nc{\p}{\partial}
\nc{\om}{\omega}
\nc{\Om}{\Omega}
\nc{\rd}{\mathrm{d}}
\nc{\Od}[1]{\mathcal{O}(#1)} 
\nc{\kp}{\kappa}

\def\({\left(}
\def\){\right)}
\def\[{\left[}
\def\]{\right]}
\def\e{\begin{equation}}
\def\q{\end{equation}}
\def\m{\begin{eqnarray}}
\def\n{\end{eqnarray}}
\nc{\Eq}[1]{Eq.~\eqref{#1}}     
\nc{\Fig}[1]{Fig.~\ref{#1}}     
\nc{\Table}[1]{Table~\ref{#1}}  
\nc{\Sec}[1]{Sec.~\ref{#1}}     
\nc{\Msun}{M_\odot}             
\nc{\fpbh}{f_{\mathrm{pbh}}}    
\nc{\fpbhn}{f_{\mathrm{pbh0}}}    
\nc{\mR}{\mathcal{R}} 
\nc{\seq}{\sigma_{\mathrm{eq}}}
\nc{\ogw}{\Omega_{\mathrm{GW}}}
\nc{\gpcyr}{\mathrm{Gpc}^{-3}\,\mathrm{yr}^{-1}}
\nc{\lvc}{LIGO/Virgo} 
\nc{\SNR}{\mathrm{SNR}} 
\nc{\mmin}{{m_{\mathrm{min}}}}
\nc{\mmax}{{m_{\mathrm{max}}}}
\nc{\Mmin}{{M_{\mathrm{min}}}}
\nc{\fmin}{{f_{\mathrm{min}}}}
\nc{\VT}{\mathrm{VT}}
\nc{\rhoGW}{\rho_{\mathrm{GW}}}
\nc{\vth}{\vec{\theta}}
\nc{\vd}{\vec{d}}
\nc{\vla}{\vec{\lambda}}
\nc{\Nobs}{N_{\mathrm{obs}}}
\nc{\av}[1]{\langle #1 \rangle} 
\nc{\km}{\mathrm{km}}
\nc{\Mpc}{\mathrm{Mpc}}
\nc{\Tobs}{T_{\mathrm{obs}}}
\nc{\Ntemp}{N_{\mathrm{temp}}}
\nc{\uni}{\mathrm{U}}
\nc{\logu}{\operatorname{\mathrm{log-U}}}
\nc{\addref}{[\textcolor{red}{add ref}] } 
\nc{\eg}{\textit{e.g.~}}
\nc{\app}{\approx}
\nc{\hf}{\frac{1}{2}}
\nc{\discuss}{\textcolor{red}{Add discussion here!}}
\nc{\red}[1]{\textcolor{red}{#1}}
\nc{\mH}{\mathcal{H}}
\nc{\cs}{c_s^2}
\nc{\Sij}[1]{S_{ij}^{(#1)}}
\nc{\vi}[1]{v_i^{(#1)}}
\nc{\no}{\nonumber}
\def\<{\left\langle}
\def\>{\right\rangle}

\nc{\bk}{\bm{k}}
\nc{\bq}{\bm{q}}
\nc{\bp}{\bm{p}}
\nc{\bl}{\bm{l}}
\nc{\bx}{\bm{x}}
\nc{\be}{\bm{\epsilon}}
\nc{\mS}{\mathcal{S}}
\nc{\te}{\tilde{\eta}}
\nc{\tp}{\tilde{p}}
\nc{\tk}{\tilde{k}}
\nc{\tx}{\tilde{x}}
\nc{\tF}{\tilde{F}}
\nc{\tA}{\tilde{A}}
\nc{\mkpq}{|\bk-\bp-\bq|}
\nc{\mpq}{|\bp-\bq|}
\nc{\mkp}{|\bk-\bp|}
\nc{\mSi}[1]{\mS^{(#1)}({\bk, \eta})}
\nc{\vk}{\vec{k}}
\nc{\kstar}{k_*}
\nc{\fstar}{f_*}
\nc{\xstar}{x_*}
\nc{\mpbh}{m_{\rm{pbh}}}
\nc{\bn}[1]{\dt\bm{t}_{\text{#1}}}
\nc{\bC}[1]{\bm{C}_{\text{#1}}}
\nc{\NTOA}{N_{\text{TOA}}}
\nc{\Nmode}{{N_{\text{mode}}}}
\nc{\ARN}{A_{\rm{RN}}}
\nc{\gRN}{\gamma_{\rm{RN}}}
\nc{\bS}{\mathbf{\Sigma}}
\nc{\br}{\mathbf{r}}
\nc{\bN}{\mathbf{R}}
\nc{\RN}{\mathrm{RN}}
\nc{\BN}{\mathrm{BN}}
\nc{\GN}{\mathrm{GN}}
\nc{\mcN}{\mathcal{N}}
\nc{\GWB}{\mathrm{GW}}
\nc{\yr}{\mathrm{yr}}
\nc{\Am}{\mathcal{A}}
\nc{\Dm}{\mathcal{D}}
\nc{\Hm}{\mathcal{H}}
\nc{\mrm}{\mathrm}
\nc{\BF}{\mathcal{BF}}
\nc{\bt}{\mathbf{t}}
\nc{\bd}{\mathbf{d}}
\nc{\ba}{\mathbf{a}}
\nc{\bnu}{\mathbf{\nu}}

\begin{document}
	
\title{Search for the Gravitational-wave Background from Cosmic Strings with the Parkes Pulsar Timing Array Second Data Release}
	
\author{Zu-Cheng Chen}
\email{zucheng.chen@bnu.edu.cn}
\affiliation{Department of Astronomy, Beijing Normal University, Beijing 100875, China}
\affiliation{Advanced Institute of Natural Sciences, Beijing Normal University, Zhuhai 519087, China}

\author{Yu-Mei Wu}
\email{Corresponding author: wuyumei@itp.ac.cn} 
\affiliation{CAS Key Laboratory of Theoretical Physics, Institute of Theoretical Physics, Chinese Academy of Sciences, Beijing 100190, China}
\affiliation{School of Physical Sciences, University of Chinese Academy of Sciences, No. 19A Yuquan Road, Beijing 100049, China}

\author{Qing-Guo Huang}
\email{Corresponding author: huangqg@itp.ac.cn}
\affiliation{CAS Key Laboratory of Theoretical Physics, 
	Institute of Theoretical Physics, Chinese Academy of Sciences,
	Beijing 100190, China}
\affiliation{School of Physical Sciences, 
	University of Chinese Academy of Sciences, 
	No. 19A Yuquan Road, Beijing 100049, China}
\affiliation{School of Fundamental Physics and Mathematical Sciences,
	Hangzhou Institute for Advanced Study, UCAS, Hangzhou 310024, China}

	
\date{\today}

\begin{abstract}
We perform a direct search for an isotropic stochastic gravitational-wave background (SGWB) produced by cosmic strings in the Parkes Pulsar Timing Array second data release. We find no evidence for such an SGWB, and therefore place $95\%$ confidence level upper limits on the cosmic string tension, $G\mu$, as a function of the reconnection probability, $p$, which can be less than 1 in the string-theory-inspired models. The upper bound on the cosmic string tension is $G\mu \lesssim 5.1 \times 10^{-10}$ for $p = 1$, which is about five orders of magnitude tighter than the bound derived from the null search of individual gravitational wave burst from cosmic string cusps in the PPTA DR2.
\end{abstract}

\maketitle

\section{Introduction}

Over the past few years, the gravitational wave (GW) community witnessed the detection of a population of GW events \cite{LIGOScientific:2018mvr,LIGOScientific:2020ibl,LIGOScientific:2021usb,LIGOScientific:2021djp} with ground-based interferometers that are sensitive in frequencies from Hz to kHz.
A pulsar timing array (PTA) \cite{1978SvA....22...36S,Detweiler:1979wn,1990ApJ...361..300F} offers a unique opportunity of extending the GW observations to the very low frequencies from nHz to $\mu$Hz, by regularly monitoring the time of arrivals (TOAs) of radio pulses from an array of highly stable millisecond pulsars in the Milky Way.
There are three major PTA projects, namely the European PTA (EPTA) \cite{Kramer:2013kea}, the North American Nanoherz Observatory for GWs (NANOGrav) \cite{McLaughlin:2013ira}, and the Parkes PTA (PPTA) \cite{Manchester:2012za}. These PTA projects have been monitoring the TOAs from dozens of pulsars with a weekly to monthly cadence for more than a decade. These PTAs along with the Indian PTA (InPTA) \cite{Joshi:2018ogr}, the Chinese PTA (CPTA) \cite{2016ASPC..502...19L} and the MeerKAT interferometer \cite{Bailes:2020qai}, support the International PTA (IPTA) \cite{Hobbs:2009yy,Manchester:2013ndt}.
Potential GW sources in the PTA frequency band include a variety of physical phenomena such as supermassive black hole binaries (SMBHBs) \cite{Jaffe:2002rt,Sesana:2008mz,Sesana:2008xk}, scalar-induced GWs \cite{Saito:2008jc,Yuan:2019udt,Yuan:2019wwo,Chen:2019xse,Yuan:2019fwv,Chen:2021nxo}, first-order phase transition \cite{Witten:1984rs,Hogan:1986qda}, and cosmic strings \cite{Lentati:2015qwp,Blanco-Pillado:2017rnf,Arzoumanian:2018saf,Yonemaru:2020bmr}. 

Recently, the NANOGrav \cite{Arzoumanian:2020vkk}, PPTA \cite{Goncharov:2021oub}, EPTA \cite{Chen:2021rqp} and IPTA \cite{Antoniadis:2022pcn} successively reported strong evidence for a stochastic common-spectrum process  modeled by a power-law spectrum in their latest data sets. However, there is no significant evidence for the tensor transverse spatial correlations, which are necessary to claim a detection of stochastic GW background (SGWB) predicted by general relativity. The origin of this spatially uncorrelated common-spectrum process (UCP) is still unknown. Further investigations indicate that the UCP can possibly come from various physical processes such as phase transitions \cite{Addazi:2020zcj,Ratzinger:2020koh,Bian:2020urb,Li:2021qer,Xue:2021gyq,NANOGrav:2021flc}, domain walls \cite{Bian:2020urb,Wang:2022rjz, Ferreira:2022zzo}, cosmic strings \cite{Blasi:2020mfx,Ellis:2020ena,Bian:2020urb,Wang:2022rjz}, or the non-tensorial polarization modes from alternative gravity theories \cite{Chen:2021wdo,NANOGrav:2021ini,Wu:2021kmd,Chen:2021ncc}.

Cosmic strings are linear topological defects that can either form in the early Universe from symmetry-breaking phase transitions at high energies \cite{Kibble:1976sj,Vilenkin:1981bx,Vilenkin:1984ib,Dvali:2003zj} or be the fundamental strings of superstring theory (or one-dimensional D-branes) stretched out to astrophysical lengths \cite{Dvali:2003zj,Copeland:2003bj}. After their formation, the intersection between cosmic strings can lead to reconnections and form loops, which will then decay due to relativistic oscillation and emit GWs.
PTA observations may detect a cosmic string network either through GW bursts emitted at cusps or through the SGWB superposed by radiation from all loops existing through cosmic history.
A null detection of the individual GW burst from cosmic strings in PPTA DR2 has been reported in \cite{Yonemaru:2020bmr}, thus placing a $95\%$ upper limit on the cosmic string tension to be $G\mu \lesssim 10^{-5}$.
In this work, we perform the first direct search for the SGWB produced from a network of cosmic strings in the PPTA DR2 data set.
The remainder of this paper is organized as follows. In \Sec{sec:CS}, we review the SGWB energy spectrum produced by the cosmic strings. In \Sec{sec:data}, we describe the data set and methodology used in our analyses. Finally, we summarize the results and give some discussion in \Sec{sec:result}.

\section{\label{sec:CS}SGWB from cosmic strings}

We now review the SGWB produced by cosmic strings following \cite{Blanco-Pillado:2017oxo}.
A cosmic string network consists of both long (or ``infinite") strings that are longer than the horizon size and loops formed from smaller strings. When two cosmic strings meet one another, they can exchange partners with a reconnection probability $p$, and form loops. Once loops are formed, they oscillate and decay through the emission of GWs \cite{Vilenkin:1981bx}, shrinking in size. A cosmic string network grows along with the cosmic expansion and evolves toward the scaling regime in which all the fundamental properties of the system scale with the cosmic time. The scaling regime can be achieved through the formation and subsequent decay of loops. The GW spectrum created by a cosmic string network is exceptionally broadband, depending on the size of the loops created.

We describe the GW energy spectrum of cosmic strings in terms of the dimensionless tension, $G \mu$, and the reconnection probability, $p$. Even though $p=1$ for classical strings, it can be less than 1 in the string-theory-inspired models. In this work, we adopt the convention that the speed of light $c=1$.
The dimensionless GW energy density parameter per logarithm frequency as the fraction of the critical energy density is \cite{Blanco-Pillado:2017oxo}
\e
\Omega_{\mathrm{gw}}(f)=\frac{8 \pi G f}{3 H_{0}^{2} p} \rho_{\mathrm{gw}}(t_0, f),
\q
where $t_0$ is cosmic time today, and $\rho_{\mathrm{gw}}$ is the GW energy density per unit frequency that can be computed by
\e
\rho_{\mathrm{gw}}(t, f)=G \mu^{2} \sum_{n=1}^{\infty} C_{n} P_{n},
\q
with
\e
C_{n}(f)=\int_{0}^{t_{0}} \frac{d t}{(1+z)^{5}} \frac{2 n}{f^{2}} \mathrm{n}(l, t).
\q
Here, $P_{n}$ is the radiation power spectrum of each loop, and $\mathrm{n}(l, t)$ is the density of loops per unit volume per unit range of loop length $l$ existing at time $t$. The SGWB from a network of cosmic strings has been computed in \cite{Blanco-Pillado:2017oxo} with a complete end-to-end method by (i) simulating the long string network to extract a representative sample of loop shapes; (ii) using a smoothing model to estimate loop shape deformations due to gravitational backreaction; (iii) computing GW spectrum for each loop; (iv) evaluating the distribution of loops over redshift by integrating over cosmological time; (v) integrating the GW spectrum of each loop over the redshift-dependent loop distribution to get the overall emission spectrum; and at last (vi) integrating the overall emission spectrum over cosmological time to get the current SGWB.
The simulations span a large parameter space of $G\mu \in [10^{-25}, 10^{-8}$], and $f \in [10^{-15}, 10^{10}]$; and the output of the expected energy density spectra has been made publicly available\footnote{\url{http://cosmos.phy.tufts.edu/cosmic-string-spectra/}}. 

\section{\label{sec:data}The data set and methodology}

\begin{table*}[!htbp]
	\begin{tabular}{c c c c}
		\hline
		\textbf{Parameter} & \textbf{Description} & \textbf{Prior} & \textbf{Comments}  \,\\
		\hline
		\multicolumn{4}{c}{White noise}\,\\	        
		$E_{k}$ & EFAC per backend/receiver system & $\uni[0, 10]$ & single pulsar analysis only \\
		$Q_{k}$[s] & EQUAD per backend/receiver system & $\logu[-8.5, -5]$ & single pulsar analysis only \\
		$J_{k}$[s] & ECORR per backend/receiver system & $\logu[-8.5, -5]$ & single pulsar analysis only \\
		\hline
		\multicolumn{4}{c}{Red noise (including SN, DM and CN)} \\
		$\Am_{\RN}$ & red noise power-law amplitude &$\logu[-20, -8]$ & one parameter per pulsar\, \\
		$\gamma_{\RN}$ &red noise power-law index  &$\uni[0,10]$ & one parameter per pulsar\, \\
		\hline
		\multicolumn{4}{c}{Band/System noise}\,\\
		$\Am_{\BN/\GN}$ & band/group-noise power-law amplitude &$\logu[-20, -8]$ & one parameter per band/system\, \\
		$\gamma_{\BN/\GN}$ &band/group-noise power-law index &$\uni[0,10]$ &one parameter per band/system\, \\
		\hline
		\multicolumn{4}{c}{Deterministic noise}\,\\
		$\Am_{\mathrm{E}}$ & exponential dip amplitude &$\logu[-10, -2]$ & one parameter per exponential dip event \, \\
		$ t_{\mathrm{E}}[\mrm{MJD}]$ &time of the event & $\uni[57050, 57150]$ for PSR J1643 & one parameter per exponential dip event \, \\
		$ $ & &$\uni[56100, 56500]$ for PSR J2145 & \, \\
		$\mrm{log_{10}} \tau_{\mrm{E}}[\mrm{MJD}]$ & relaxation time for the dip &$\uni[\mrm{log_{10}}5, 2]$ &one parameter per exponential-dip event \, \\
		$\Am_{\mathrm{G}}$ & Gaussian bump amplitude &$\logu[-6, -1]$ & one parameter per Gaussian bump event \, \\
		$ t_{\mrm{G}}[\mrm{MJD}]$ & time of the bump &$\uni[53710, 54070]$ &one parameter per Gaussian bump event \, \\
		$ \sigma_{\mrm{G}}[\mrm{MJD}]$ & width of the bump&$\uni[20, 140]$ &one parameter per Gaussian bump event \, \\
		$\Am_{\mathrm{Y}}$ & annual variation amplitude &$\logu[-10, -2]$ & one parameter per annual event \, \\
		$ \phi_{\mrm{Y}}$ & phase of the annual variation &$\uni[0, 2\pi]$ &one parameter per annual event \, \\
		\hline
		\multicolumn{4}{c}{SGWB from cosmic string}\,\\
		$G\mu$ & cosmic string tension & $\logu[-15, -8]$ & one parameter per PTA \\
		$p$ & reconnection probability & $\logu[-3, 0]$ & one parameter per PTA \\
		\hline
	\end{tabular}
	\caption{Model parameters and their prior distributions used in the Bayesian inference.}
	\label{tab:prior}
\end{table*}

The PPTA DR2 \cite{Kerr:2020qdo} data set includes pulse TOAs from high-precision timing observations for 26 pulsars collected with the 64-m Parkes radio telescope in Australia. The data were acquired between 2003 and 2018, spanning about $15$ years, with observations taken at a cadence of approximately three weeks \cite{Kerr:2020qdo}. Details of the observing systems and data processing procedures are described in \cite{Manchester:2012za,Kerr:2020qdo}.

To search for the GW signal from the PTA data, one needs to provide a comprehensive description of the stochastic processes that can induce the arrival time variations. The stochastic processes can be categorized as being correlated (red) or uncorrelated (white) in time. A careful analysis of the noise processes for individual pulsars in the PPTA sample has been performed in \cite{Goncharov:2020krd}, showing that the PPTA data sets contain a wide variety of noise processes, including instrument dependent or band-dependent processes. Similar to \cite{Wu:2021kmd}, we adopt the noise model developed in \cite{Goncharov:2020krd} to characterize the noise processes. After subtracting the timing model of the pulsar from the TOAs, the timing residuals $\dt \bt$ for each single pulsar can be decomposed into (see \eg \cite{Lentati:2016ygu})
\e
\dt\bm{t} = M \bm{\epsilon} + \bn{RN} + \bn{DET} + \bn{WN} + \bn{CP}.
\label{dt}
\q
The first term $M \be$ in the above equation accounts for the inaccuracies in the subtraction of timing model \cite{Chamberlin:2014ria}, where $M$ is the timing model design matrix obtained from \texttt{TEMPO2} \cite{Hobbs:2006cd,Edwards:2006zg} through \texttt{libstempo}\footnote{\url{https://vallis.github.io/libstempo}} interface, and $\bm{\epsilon}$ is a small offset vector denoting the difference between the true parameters and the estimated parameters of timing model. 
The second term $\bn{RN}$ is the stochastic contribution from red noise \cite{Goncharov:2020krd}, including achromatic spin noise (SN) \cite{Shannon:2010bv}; frequency-dependent dispersion measure (DM) noise \cite{Keith:2012ht}; frequency-dependent chromatic noise (CN) \cite{Andrew:cn}; achromatic band noise (BN) and system (``group”) noise (GN) \cite{Lentati:2016ygu}. We use $30$ frequency components for the red noise of the individual pulsar.
The third term $\bn{DET}$ represents deterministic noise \cite{Goncharov:2020krd}, including chromatic exponential dips \cite{Lentati:2016ygu}, extreme scattering events \cite{Keith:2012ht}, and annual dispersion measure variations \cite{Coles:2015uia}.
The fourth term $\bn{WN}$ represents white noise (WN), including a scale parameter on the TOA uncertainties (EFAC), an added variance (EQUAD), and a per-epoch variance (ECORR) for each backend/receiver system \cite{NANOGrav:2015aud}.
The last term $\bn{CP}$ is the stochastic contribution due to the common-spectrum process (such as an SGWB), which is described by the cross-power spectral density \cite{Thrane:2013oya}
\e
S_{I J}(f)=\frac{H_{0}^{2}}{16 \pi^{4} f^{5}} \Gamma_{I J}(f)\, \Omega_{\mathrm{GW}}(f),
\q
where $\Gamma_{I J}$ is the Hellings \& Downs coefficients \cite{Hellings:1983fr} measuring the spatial correlations of the pulsars $I$ and $J$. Following \cite{NANOGrav:2020bcs}, we use 5 frequency components for the common process among all of the pulsars.

We perform the Bayesian parameter inferences based on the methodology in \cite{NANOGRAV:2018hou,NANOGrav:2020bcs} to search for the SGWB from cosmic strings in the PPTA DR2 data set. Since it is challenging to obtain a complete noise model for pulsar J0437$-$4715 and pulsar J1713+0747 \cite{Goncharov:2020krd}, we do not include these two pulsars. A summary of the noise model for the $24$ pulsars used in our analyses can be found in Table~1 of \cite{Wu:2021kmd}. We quantify the model selection scores by the Bayes factor defined as
\e
\BF \equiv \frac{\rm{Pr}(\mathcal{D}|\mathcal{M}_2)}{\rm{Pr}(\mathcal{D}|\mathcal{M}_1)},
\q
where $\rm{Pr}(\mathcal{D}|\mathcal{M})$ measures the evidence that the data $\mathcal{D}$ are produced under the hypothesis of model $\mathcal{M}$. Model $\mathcal{M}_2$ is preferred over $\mathcal{M}_1$ if the Bayes factor is sufficiently large. As a rule of thumb, $\BF \le 3$ implies the evidence supporting the model $\mathcal{M}_2$ over $\mathcal{M}_1$ is ``not worth more than a bare mention" \cite{BF}. In practice, we estimate the Bayes factor using the \textit{product-space method} \cite{10.2307/2346151,10.2307/1391010,Hee:2015eba,Taylor:2020zpk}.
Our analyses are based on the latest JPL solar system ephemeris (SSE) DE438 \cite{DE438}. We first infer the parameters of each single pulsar without including the common-spectrum process $\bn{CP}$ in \Eq{dt}, and then fix the white noise parameters to their max likelihood values from single-pulsar analysis to reduce the computational costs as was commonly done in literature (see \eg \cite{NANOGRAV:2018hou,NANOGrav:2020bcs}). We use the open-source software packages \texttt{enterprise} \cite{enterprise} and \texttt{enterprise\_extension} \cite{enterprise_extensioins} to calculate the likelihood and Bayes factors and use \texttt{PTMCMCSampler} \cite{justin_ellis_2017_1037579} package to do the Markov chain Monte Carlo sampling. Similar to \cite{Aggarwal:2018mgp,NANOGrav:2020bcs}, we use draws from empirical distributions based on the posteriors obtained from the single-pulsar Bayesian analysis to sample the parameters of red noise and deterministic noise.
Using empirical distributions can reduce the number of samples needed for the chains to burn in.
The model parameters and their prior distributions are listed in \Table{tab:prior}.

\section{\label{sec:result}Results and discussion}

\begin{figure}[htbp!]
	\centering
	\includegraphics[width=0.5\textwidth]{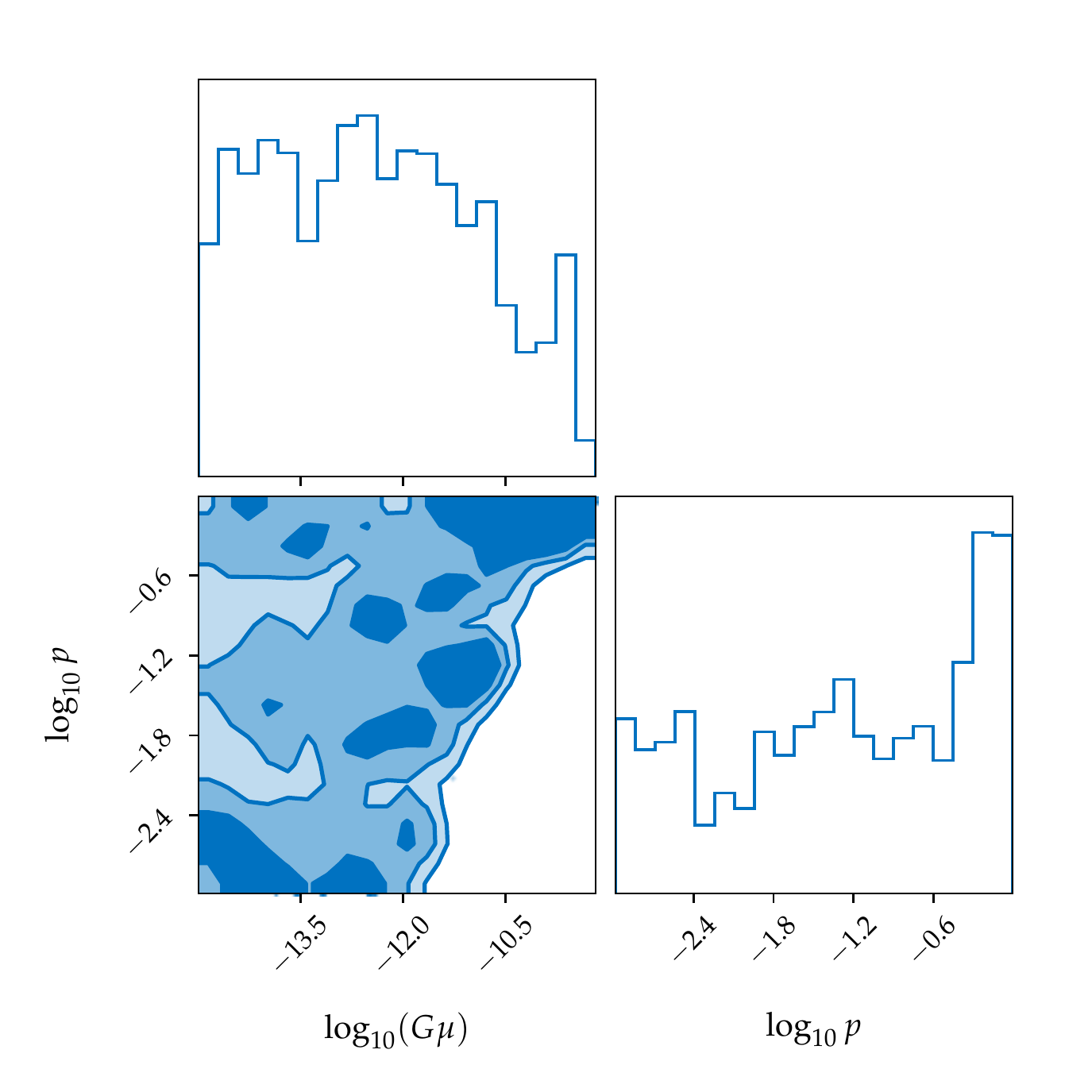}
	\caption{\label{p_Gmu_post} One- and two-dimensional marginalized posterior distributions for the cosmic string tension, $G\mu$, and the reconnection probability, $p$. We show both the $1 \sigma$ and $2 \sigma$ contours in the two-dimensional plot.}
\end{figure}

\begin{figure}[htbp!]
	\centering
	\includegraphics[width=0.5\textwidth]{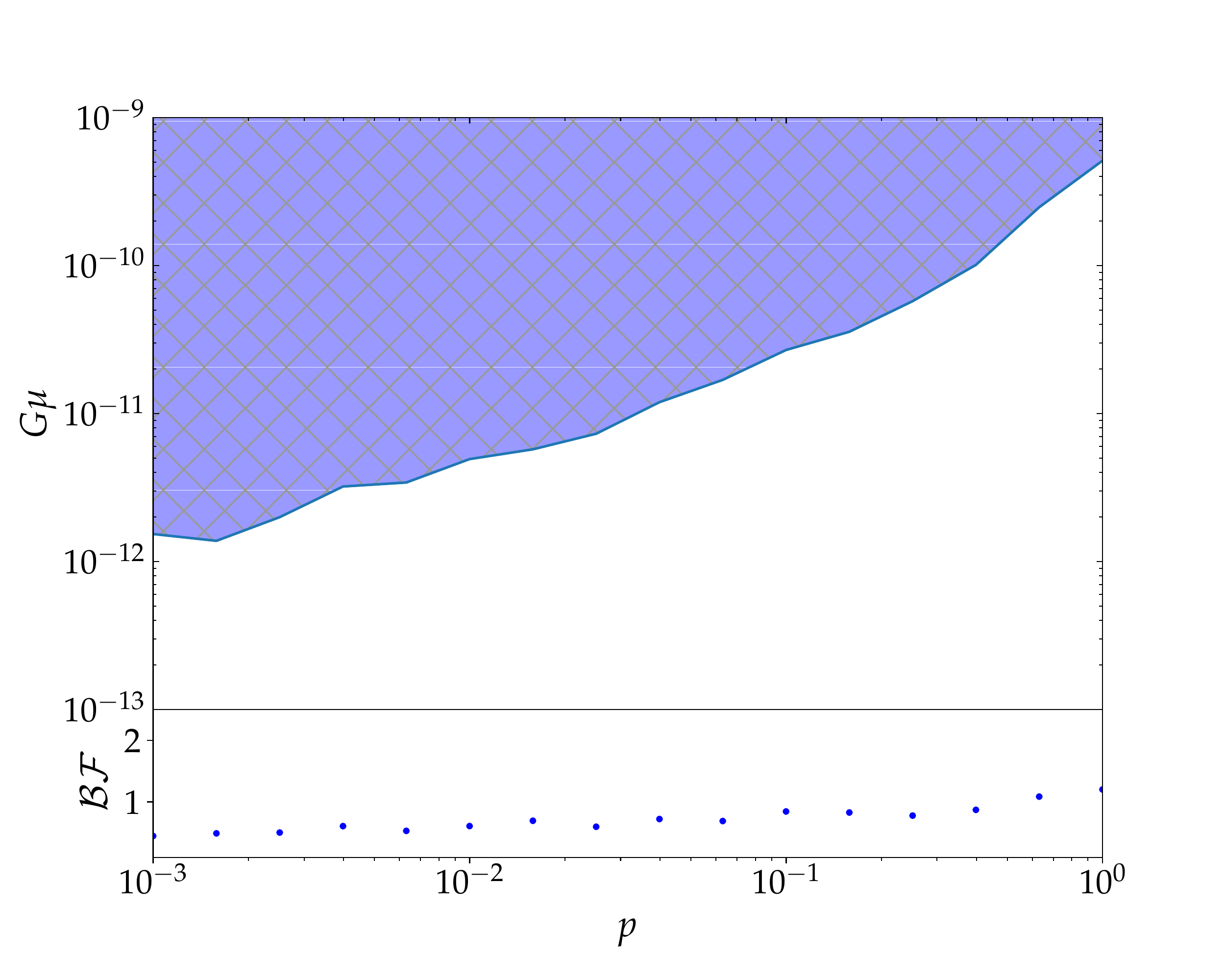}
	\caption{\label{p_Gmu_upperlimit} \textbf{Top panel}: the $95\%$ CL upper limits on the cosmic string tension, $G\mu$, as a function of the reconnection probability, $p$, from the PPTA DR2. \textbf{Bottom panel}: the corresponding Bayes factor, $\BF$, as a function of the reconnection probability, $p$.
}
\end{figure}

We first consider a model in which both the cosmic string tension $G\mu$ and the reconnection probability $p$ are free parameters. 
\Fig{p_Gmu_post} shows the posterior distributions of the $G\mu$ and $p$ parameters obtained from the Bayesian search. The Bayes factor of the model including both the UCP and CS signal versus the model including only the UCP is $\BF^{\mrm{UCP+CS}}_{\mrm{UCP}} = 0.591\pm 0.008$, indicating no evidence for an SGWB signal produced by the cosmic string in the PPTA DR2. 

We also consider models in which the reconnection probability $p$ is fixed to a specific value while the cosmic string tension $G\mu$ is allowed to vary. The lower panel of \Fig{p_Gmu_upperlimit} shows the Bayes factor as a function of reconnection probability. For all the values of $p \in [10^{-3}, 1]$, we have $\BF^{\mrm{UCP+CS}}_{\mrm{UCP}} \lesssim 3$, confirming that there is no evidence for an SGWB produced by cosmic strings in the PPTA DR2. We, therefore, place $95\%$ confidence level upper limit on cosmic string tension $G\mu$ as a function of reconnection probability $p$ as shown in \Fig{p_Gmu_upperlimit}. The blue shaded region indicates parameter space that is excluded by the PPTA DR2. For $p=1$, the upper bound on the cosmic string tension is $G\mu \lesssim 5.1 \times 10^{-10}$, which is about five orders of magnitude tighter than the bound of $G\mu \lesssim 10^{-5}$ \cite{Yonemaru:2020bmr} derived from the null search of individual gravitational wave burst from cosmic string cusps in the PPTA DR2. Note that $\Omega_{\mathrm{gw}}$ is enhanced for $p<1$, and therefore tighter constraints on $G\mu$ are obtained for $p<1$. 

To sum up, we have searched for the SGWB produced by a cosmic string network in the PPTA DR2 in the work. We find no evidence for such SGWB signal, and therefore place $95\%$ upper limit on cosmic string tension as a function of reconnection probability.

\begin{acknowledgments}
	We thank Xing-Jiang Zhu, Zhi-Qiang You, Xiao-Jin Liu, Zhu Yi, and Shen-Shi Du for useful discussions.
	We acknowledge the use of HPC Cluster of ITP-CAS and HPC Cluster of Tianhe II in National Supercomputing Center in Guangzhou. This work is supported by the National Key Research and Development Program of China Grant No.2020YFC2201502, grants from NSFC (grant No. 11975019, 11991052, 12047503), Key Research Program of Frontier Sciences, CAS, Grant NO. ZDBS-LY-7009, CAS Project for Young Scientists in Basic Research YSBR-006, the Key Research Program of the Chinese Academy of Sciences (Grant NO. XDPB15).
\end{acknowledgments}
\bibliography{./ref}

\end{document}